# Application of image processing in optical method, moiré deflectometry for investigating the optical properties of zinc oxide nanoparticle

Fatemeh Jamal[a,b], Fatemeh Ahmadi[b,c,*], Mohammad Khanzadeh [a,b], Saber Malekzadeh[d]

[a] Department of Physics, Faculty of Sciences, Vali-e-Asr University of Rafsanjan, 77188-97111 Rafsanjan, Iran

[b] Advanced Optics Lab, Vali-e-Asr University, 77188-97111 Rafsanjan, Iran

[c] Department of Physics, Shahreza Branch, Islamic Azad University, 311-86145 Shahreza, Iran

[d] Department of Computer Science, University of Tabriz, 91422-32172 Tabriz, Iran

**Abstract**

Nowadays for measurement of refractive index of nanomaterials usually spectro-photometric and mechanical methods are used which are expensive and indirect. In this paper for measuring these parameters of zinc oxide nanomaterial with two different stabilizers, a simple optical method, Moiré deflectometry, which is based on wave front analysis and geometric optics is used. In the Moiré deflectometry method, the beam of light from the laser diode passes through the sample. As a result of that, a change in the sample environment is observed as deflections of the fringes. By recording of these deflections using CCD and image processing with MATLAB, the nanomaterials refractive indices can be calculated. Due to the high accuracy of this method and improved the image processing code in Matlab, the smallest changes of the refractive index in the sample can be measured. Digital Image processing is used for processing images in a way that features can be selected and being shown. The results obtained in this method show a good improvement over the other used methods.

**1-introduction**

Nonlinear optics is a branch of modern optics which has more significance in comparison with laser physics. It also describes the behavior of light in the nonlinear material. In recent researches on nonlinear optics, the new effects with the process of interaction laser radiation with different materials have been observed.

The interaction between the laser and optical material can modify the optical properties in the material so that the laser source, on the other hand, can sufficiently provide high light intensities to modify the optical properties of materials. Normally, when laser light is enough strong, can produce nonlinear optical phenomena [1].

[*] Corresponding author.
 E-mail address: Fatemeh.ahmadi1991@yahoo.com (F. Ahmadi).

Nonlinear optics remained unexplored until the discovery in 1961 of second-harmonic generation by Peter Franken et al, shortly after the construction of the first laser by Theodore Harold Maiman. However, some nonlinear effects were discovered before the development of the laser. In nonlinear optical effect, we can name two important works like Pockels and Kerr effects. The Kerr effect is a change in the refractive index of materials in response to an applied electric field and the Pockels effect Makes change birefringence in an optical medium by an electric field. In this regard, the development of the nanotechnology provided new opportunities for the study of the nonlinear optics. Nonlinear optics used widely in many fields such as laser technology, optical communication, data processing, and optical computing. It seems that the measurement of nonlinear optical parameters of nanomaterials have drawn a lot of attention because of their fast response and high level of nonlinearity so can be used in different optical methods for investigating that optical parameter, therefore the Nanomaterials with large nonlinear responses, can be used as comparisons in nonlinear microscopes, photocatalysts, and light- limited applications [1].

In the recent years, the linear optical properties of nanomaterials have been extensively investigated and the potential of these materials for nonlinear optical applications has been largely untapped. These materials can be used in new industries such as laser technology, pulse laser deposition and optical constraints. This material plays a key role in the development of many optical technologies and can be used to process optical signal information at high speeds. So it is very important to know the optical properties of nanomaterials, especially their nonlinear refractive index and absorption coefficient [2].

Many techniques have been developed to characterize the nonlinear refractive index and absorption coefficient of materials. One of the optical methods used to measure nonlinear properties is moiré deflectometry. Through this method, it is possible to calculate the nonlinear refractive index of materials by ray deflection which is very applicable in a variety of fields in science and engineering. One of the remarkable cases is the measurement of the dimensions and knowing of the internal structure topography and external tomography of materials and objects which are some types of out-of-plane moiré method [3-4].

Moiré deflectometry as a simple optical method is used for nanomaterials characterization. This method is based on wave front analysis and geometric optics. In moiré deflectometry, the incident beam wavelength, ambient pressure, and humidity are important elements in measuring the refractive index and absorption coefficient in the nanomaterials. For instance, Farahani et al. in 2017 used the moiré deflectometry method to measure the nonlinear refractive index and absorption coefficient of dissolved graphene oxide nanoparticles in water [5-6]. In this research, they used an Nd: YAG laser with the second harmonic (532 nm wavelength) as the pump laser and 5-mw neon-helium laser as the prop laser.

Two similar identical Ranchi grating $G_1$ and $G_2$ with a pitch of 0/1 mm, and 6 cm distance between them are used. In this article, the theory of moiré deflectometry for measuring nonlinear refractive index and absorption coefficient is studied [5].

In this study, the nonlinear effect such as a refractive index of zinc oxide nanoparticles was investigated using an experimental optical method, Moiré deflectometry. We describe the experimental details and present theoretical analysis for obtaining the sign of nonlinear refractive index of materials, $n_2$, using the moiré fringe patterns. Furthermore, we studied the refractive index distribution for these kinds of nanomaterials and by using the deflection on Moiré fringes and refractive indices, we investigated the absorption coefficient.

Detection of fringes deflections has also been used to accurately measuring and observing the slightest changes in digital image processing. Of the main advantage of this method is that by processing the images recorded with the CCD, the specification of the various parameters is obtained with a measurement, but in the other method such as Z-scan method, only the point data is provided in each of measurements. Our results are consistent with the results obtained for this type of nanomaterials using the other methods such as Z-scan [7].

## 2. Theoretical and Experimental Moiré deflectometry:

In a usual configuration of Moiré deflectometer or tomography consisting of two similar Ranchi gratings, the setup is sensitive to the incoming wave-front changes only is one direction normal to its rolling. in this work, we have used to rotational Moiré deflectometry method that way the first grating $G_1$ was fixed and those in the second grating rotate in the z-direction, clockwise or anticlockwise, as shown an fig.1.

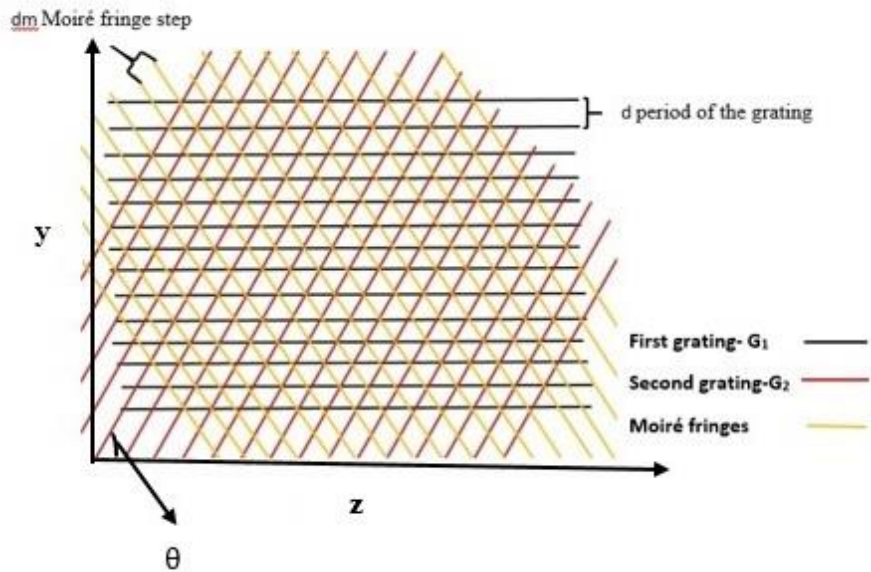

**Figure 1**: Formation of rotation Moiré fringes by using the two similar Ranchi grating

In Fig.1, d is the period of the Ranchi gratings $G_1$ and $G_2$ with a pitch of 0.2 mm/line. The changes in the Moiré patterns are related to grating separation in the experimental setup so in this experiment the distance between planes of is 280 mm which is the fourth Talbot distances. The Talbot distance is characterized by $Z_k=2d^2/\lambda$, where, d is the period of the grating, $\lambda$ is the wavelength of light. The setup which we use in this study is same as setup in previous work [11], that observed in figure.2.

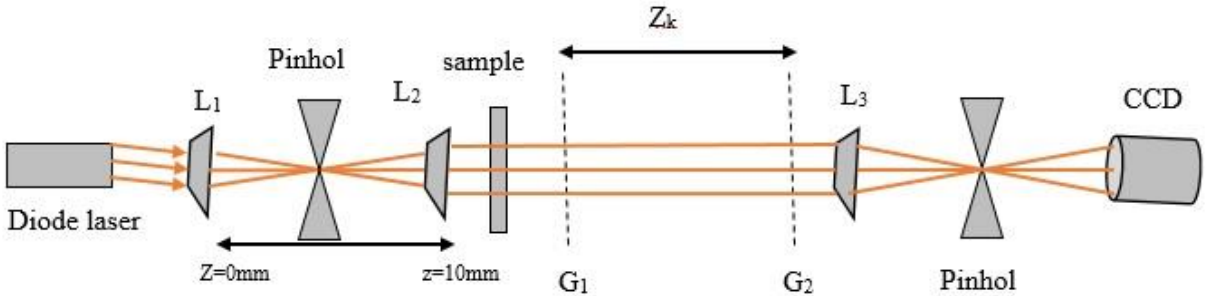

**Figure 2**: Schematic setup for nanomaterial. $L_1$ and $L_2$ are the focal lenses, $L_3$ is the parallel lens, $G_1$ fixed grating, $G_2$ rotate grating, and CCD for record the Moiré image

According to the fig. 2, The collimated beam passes through the sample which in this experiment is zinc oxide. By passing the laser beam through the grating $G_1$ and by rotate the grating $G_2$ the size of the angle θ, the lines of the image on the grating $G_2$ change of the required size dδ, the deflected laser beam due to changes in the refractive index of the sample yields the shifted self-images of the gratings, and resultant Moiré fringes show rotation, according to this Moiré fringes displaced dmδ in every place. [8-9 and 10].

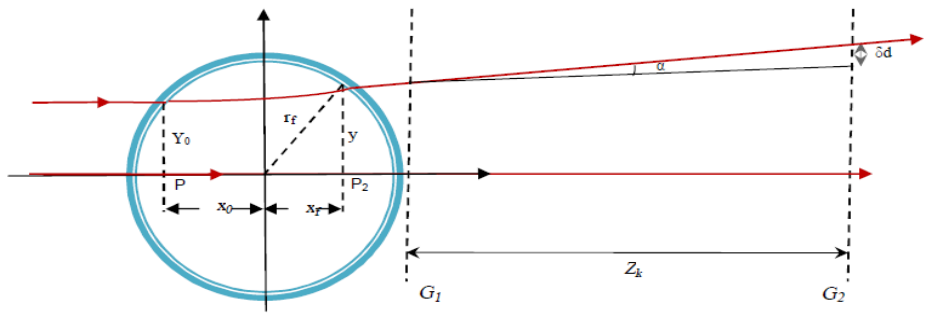

**Figure 3**: Light beam deflection, passing through phase object and grating G1 and G2 with phase object [11].

As schematically shown in Fig. 2, the value of beams of angle deflection can be calculated as follows:

$$\alpha(y, z) = \frac{\delta d}{Z_k} = \frac{d}{Z_k} \frac{\delta dm(y, z)}{dm} \quad (1)$$

In this equation, d, dm, $Z_k$ and α, are grating step for $G_1$ and $G_2$, Moiré fringe step, Talbot distance, and ray deflection angle, respectively. Therefore, the changes of refractive index can be calculated as follows.

$$n(\rho, z) - n_f = \frac{n_f}{\pi} \int_r^{r_f} \frac{\alpha(y, z)}{\sqrt{y^2 - r^2}} dy \quad (3)$$

According to this equation n (r, z) is the sample of refractive index and $n_f$ is the refractive index of the air. We can solve the integral by using image processing and Matlab programming and obtain the refractive index of the zinc oxide nanoparticles n (r, z) [11-12 and 13].

In order to measure $\beta$, refractive index change is calculated for two different fringes (distance), $Z_1$ and $Z_2$ that $\Delta n(0, Z_1) = n_2 I_0 e^{-A z_1}$ and $\Delta n(0, Z_2) = n_2 I_0 e^{-A z_2}$, respectively. So we can obtain the absorption coefficient for each pair of neighbour fringes, as follows[5]:

$$\beta = \frac{1}{z_2 - z_1} \ln\left(\frac{\Delta n(0, z_1)}{\Delta n(0, z_2)}\right) \quad (4)$$

In this research, the refractive index and absorption coefficient can be calculated for each point of Moiré fringes which these parameters are comparable [5-13].

Due to the interaction of laser beam with zinc oxide at the cell, we observe the different deflection all over the nanomaterial, therefore, it can be investigated the relation between the refractive index and absorption coefficient with the length of the cell [5].

### 3- Processing the images of Moiré deflectometry

In the previous work [11], we used only the steady effect of plasma jet on Moiré fringes, but in this study we investigated the intensity of laser prop on zinc oxide nanomaterials. Due to the zinc oxide solution is colloidal, we observed the different deflection in the patterns which show that the nanomaterial is non-steady. also in this study we have used an improved algorithm for processing the Moiré fringes to investigated this differences which we can calculate the refractive index and absorption coefficient for all over the fringes (fig.4).

Digital Image processing based is the ability to process images in a way that features can be selected and shown. Features like how the lines are curved or deviated. In this paper image processing algorithms are used to finding the amount of deviation of laser lines when they are passing. This measurement is the best way when high-resolution cameras are accessible. So in high-resolution images, the amount of deviation of laser lines are better calculated than any measurement algorithm with the human eye.

The most important part in calculating the refractive index of nanomaterial is the image processing, with placing the sample in the set-up, images recorded with the CCD at two states, the first reference image and second deflected image. For investigating the deflection in the image processing part, first images of Moiré deflectometry are given as input to deflection measuring algorithm. Then every image changed to black and white images by a suitable threshold to bold passing fringes and make them different from the background. Then every passing fringe is made thinner by getting average of its high and low borders, the maximum deflection and its deviated way of passing fringes is calculated to know the maximum deflection of fringes when they pass from the nanomaterial. All calculations were based on one laser line based on its way along the time and crossing position, and the 2d diagram is drawn based on both deflection and position [14-15].

The important point is the calibration of CCD. In this research the size of images is calculated in pixels, therefore to measure the amount of deflection accurately, measurements must be converted to millimeters or centimeter. To do this the cell with 1 mm length is used. At last, by using these measurements the refractive index of nanomaterial can be calculated.

**4. Result and discussion**

In this experiment, the optical properties of zinc oxide nanoparticles with a ZnO chemical formula were investigated. The size of zinc oxide nanoparticles is between 10-30 nm. For the measurement of the refractive index Ethanol was added to obtain a clear solution which we can observe the changes in the Moiré fringes patterns. Ethanol is often used as a solvent for zinc oxide. Although the zinc oxide nanoparticles were generally colloidal and insoluble we can use the stabilizers to turn it into a clear solution for some time. So at first, zinc oxide nanoparticle was dissolved in 21 ml of 68% ethanol. than 15 drops of Monoethanolamine was added as the stabilizer and transparency to this solution. The resulting solution was stirred for 11 hours by using the magnetic stirred or ultrasonic.

In this research in order to measure the optical properties of samples, Moiré deflectometry technique was employed. The samples were prepared and poured into a quartz cell with a thickness of 3 mm. Fig.1 shows the layout of Moiré deflectometry experimental apparatus. As it is observed the sample is placed between the second lens $L_2$ and first grating G1. The laser beam then passes through the nanoparticles in quartz cell, then through grating $G_1$ and $G_2$ and forms Moiré fringe patterns on a CCD camera through lens $L_3$ and is recorded by a computer.

CCD records reference fringes (Fig.4a) and deflected fringes (Fig.4b). As it is observable in Fig.3, image A is the reference without sample and image B is the Moiré fringe after placing the sample, which according to it, the maximum of deflection is created in the first Moiré fringe.

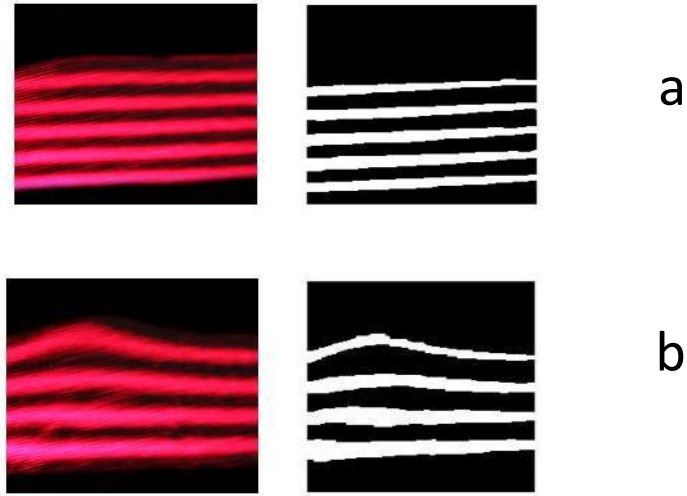

**Figure 4**: a) reference fringes b) deflected fringes in the zinc oxide nanoparticle

By applying image processing in MATLAB, explained in the previous section, according to the equation 1, the Moiré fringe deflection versus the Moiré fringe spacing $\frac{\delta dm(y,z)}{dm}$ is obtained in each point of the fringe, which is shown in Fig.5. In this images, by using the algorithms to find the position of the fringes, the references and deflected fringes position were found and plot them. As shown in fig.5c, $\delta dm$ is the difference between the reference fringe position and the deflected fringe position.

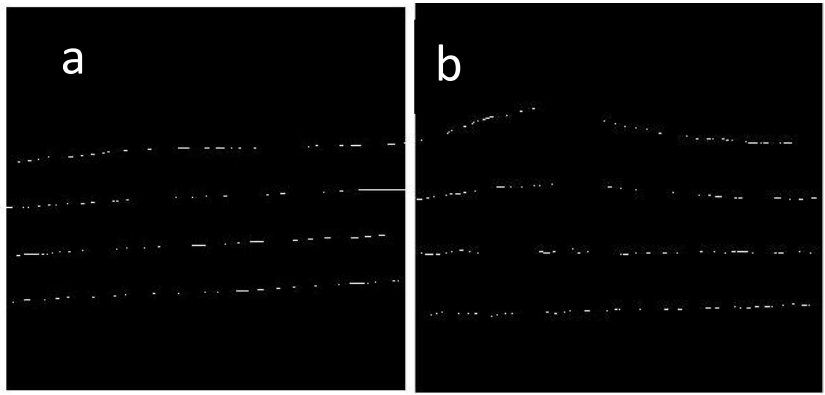

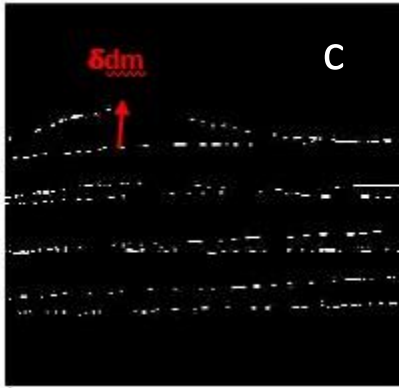

**Figure 5**: position of Moiré fringes by image processing (a) reference (b) deflected Moiré fringes (c) **δ**dm

According to fig.6, it is observed that Moiré fringes deflection in terms of distance from the center of cell. Noting that this study and according to the fig.6, due to the different intensity all over the quartz cell, the observed deflections in the different place of the cell are differences.

According to the beam deflection diagram in fig.6, it is observed that the first Moiré fringes deflection is larger than the other fringes, and therefore, according to Eq.2, the larger the deflection of Moiré fringes, the bigger the changes in the refractive index distribution. Figure 7 shows the variation of the refractive index distribution for zinc oxide nanoparticle all over the deflection.

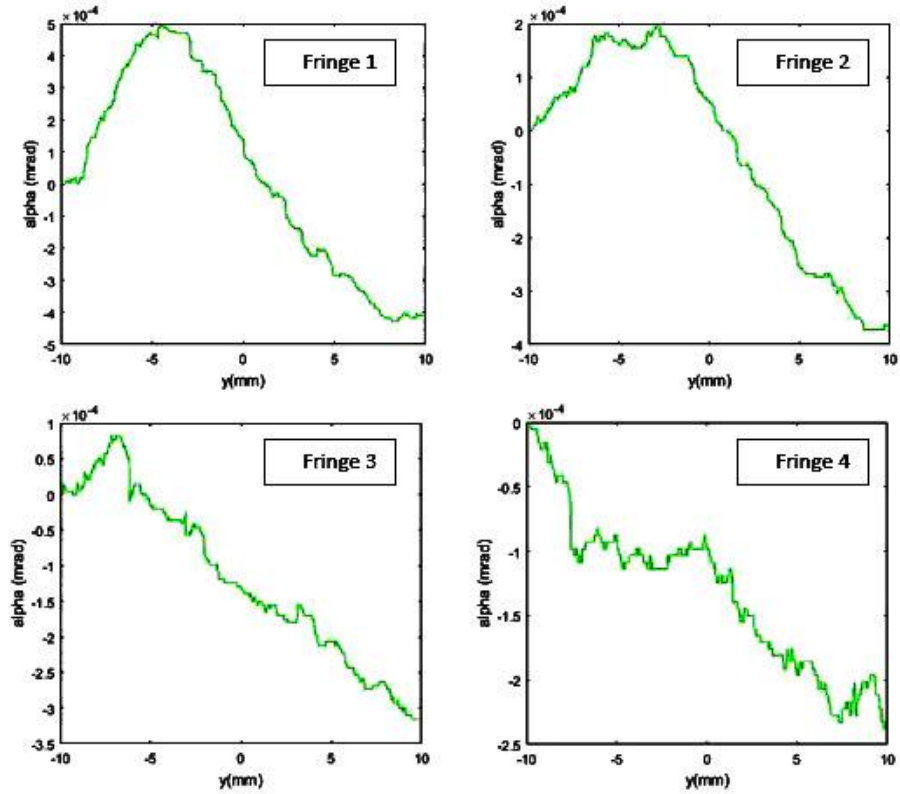

**Figure 6**: the deflecttion of 4 moire fringes in zinc oxide nanoparticle

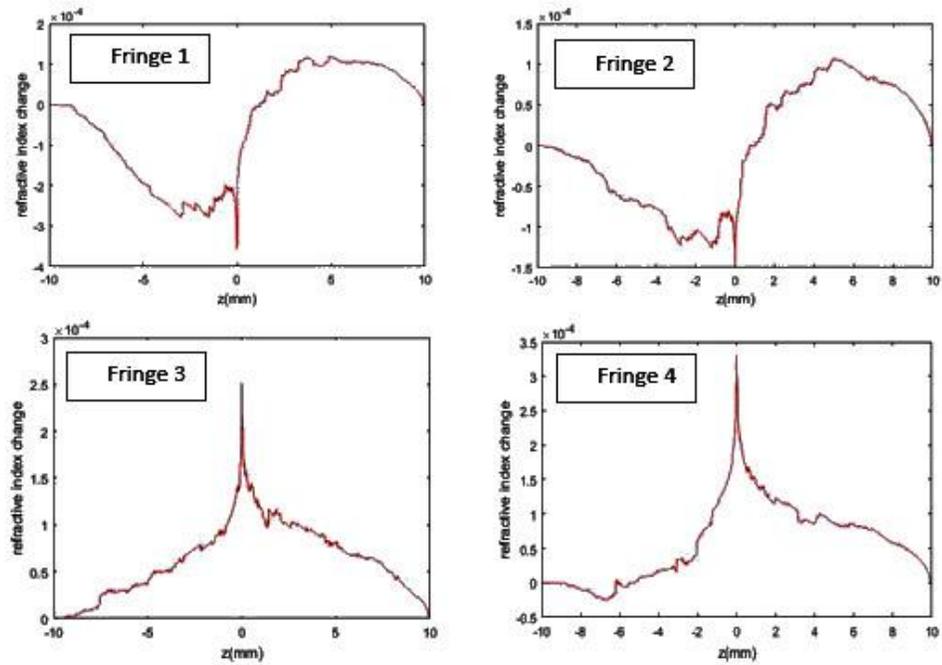

**Figure 7**: the refractive index distribution of zinc oxide versus distance from the nanoparticle canter

According to Fig.7, we calculate the refractive index distribution for all fringes and we could investigate the absorption coefficient nonlinear of zinc oxide nanoparticle. For example, in first fringe which have the higher deflection the changes of refractive index are $\Delta n = n_2 - n_f = -2.8 \times 10^{-4}$ and so $n_2 = -1.983 \times 10^{-4}$, then by using the equation 4 we could investigate the absorption coefficient nonlinear of zinc oxide nanoparticle for this fringe is the $\beta = 2.28$ cm$^{-1}$.

Based on the fig.6 and fig.7 we could investigate the refractive index and absorption coefficient to all over the Moiré deflection which it shows the high accuracy in image processing and Moiré deflectometry.

The point to be considered in this study is the use of a suitable stabilizer for making zinc oxide solution to maintain a good stability during the test and due to the colloidal do not to be precipitate quickly. As we have already stated, our selective stabilizer is mono-ethanolamine, which we can measure accurately the optical parameters by using the recorded images. The second stabilizer we used was DI-ethanolamine, which, according to the obtained images (Fig. 8), had negative effects on the optical properties of the zinc oxide nanoparticle, so that by placing it in the Moiré setup, there are no changes in the Moiré fringes.

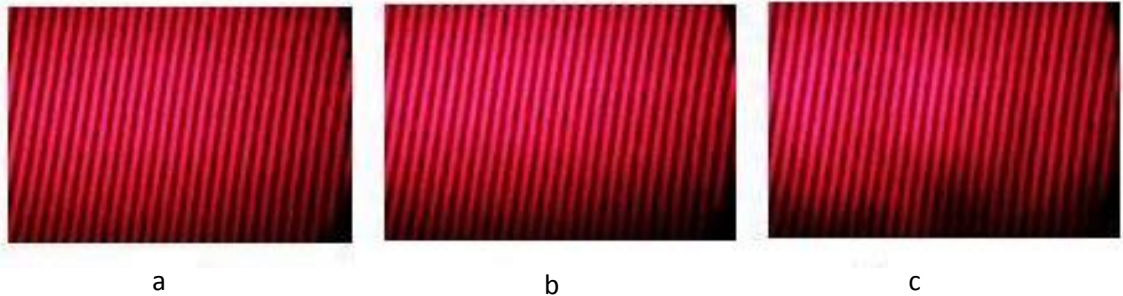

a      b      c

**Figure 8**: Moiré fringes with Di-ethanolamine a) empty cell b) cell with nanomaterial c) influence by laser

Therefore, considering these studies and measurements and comparing the results with other experiments carried out on this nanoparticle, and according to images (fig.4, fig.8), it can be concluded that the best type of stabilizer that maintains the optical properties of the nanoparticle zinc oxide is the stabilization of mono-ethanolamine.

## 5. Conclusion

In this research, zinc oxide refractive index distribution has been studied by Moiré deflectometry optical method and image processing. Due to the sensitivity and precision of this optical method, small variations in the refractive index distribution of the sample was measured. These variations are very small so they are not measurable by other methods. The proposed Moiré deflectometry technique is a non-scanning method, and the sign of refractive index can be achieved immediately and real time.

Also by using the image processing, the investigation of smallest changes and deflections in moiré fringes can be done. By observing the deflection of Moiré fringes, the sign of nonlinear refractive index can be determined. This method is simple, fast, and not sensitive to environmental noise and vibrations. Also, it does not need calibration or analysis of fringe pattern. Results show that between four Moiré fringes (Fig.5), the first fringe deflection is the highest in comparison to the other fringes, so we can use them in image processing to probe the refractive index of zinc oxide nanoparticle. The results show that with decreasing the deflection along the fringes, the changes of refractive index are decreases, too. Another hand by decreasing the changes of refractive index, the nonlinear absorption are increases.

**References**

[1] R. W. Boyd," Nonlinear optics, Handbook of Laser Technology and Applications". (2003). 161-183.

[2] Lin, Su-Shia. "Optical properties of TiO2 nanoceramic films as a function of N–Al co-doping." Ceramics International 35.7 (2009): 2693-2698.

[3] Kafri, O., & Livnat, A. (1981). Reflective surface analysis using moiré deflectometry. Applied Optics, 20(18), 3098-3100.

[4] Glatt, I., & Kafri, O. (1988). Moiré deflectometry—ray tracing interferometry. Optics and lasers in engineering, 8(3-4), 277-320.

[5] Farahani, S. S., & Madanipour, K. (2017, June). Nonlinear absorption coefficient measurement of nanofluids using Moire deflectometry technique. In Optical Methods for Inspection, Characterization, and Imaging of Biomaterials III (Vol. 10333, p. 103331I). International Society for Optics and Photonics.

[6] I. Amidror, I. Amidror, I. Amidror, and I. Amidror, The Theory of the Moir_e Phenomenon (Kluwer Academic, Dordrecht, Boston).

[7] Irimpan, L., Nampoori, V. P. N., Radhakrishnan, P., Krishnan, B., & Deepthy, A. (2008). Size-dependent enhancement of nonlinear optical properties in nanocolloids of ZnO. Journal of applied physics, 103(3), 033105.

[8] Ara, M. M., Mousavi, S. H., Salmani, S., & Koushki, E. (2008). Measurement of nonlinear refraction of dyes doped liquid crystal using moiré deflectometry. Journal of Molecular Liquids, 140(1-3), 21-24.

[9] Y. Song, Y. Y. Chen, A. He, and Z. Zhao, "Theoretical analysis for moir_e deflectometry from diffraction theory," JOSA A 26(4), 882–889 (2009).

[10] C. Shakher and A. K. Nirala, "A review on refractive index and temperature profile measurements using laser-based interferometric techniques," Opt. Laser Eng. 31(6), 455–491 (1999).

[11] Khanzadeh, M., Jamal, F., & Shariat, M. (2018). Experimental investigation of gas flow rate and electric field effect on

refractive index and electron density distribution of cold
atmospheric pressure-plasma by optical method, Moiré deflectometry. Physics of Plasmas, 25(4), 043516.
[12] I.Amidror,*"The Theory of the Moiré Phenomenon"*,Volume I: Periodic Layers Second Editon
[13] R.F.Anastasi, "an introduction to moire methods with applications in composite matrials", U.S. Army matrerials technology laboratory, Massachusetts 02172-0001 August( 1992)
[14] Vernon, D. (1991). Machine Vision Prentice-Hall International.
[15] Gonzales, R. C., & Woods, R. E. (1992). Digital Image Processing, Addison & Wesley Publishing Company. Reading, MA.